\documentclass[10pt,letterpaper]{article}
\usepackage{opex3}
\usepackage{color}
\usepackage{threeparttable}
\usepackage{epstopdf}
\usepackage{graphicx}
\usepackage{mathrsfs}
\usepackage{amssymb}
\usepackage{amsmath}
\usepackage{dsfont}
\usepackage{cite}

\begin{document}

\title{Quantum state engineering by periodical two-step modulation in atomic system}

\author{Zhi-Cheng Shi$^{1,2}$, Du Ran$^{1,2}$, Li-Tuo Shen$^{1,2}$, Yan Xia$^{1,2,*}$, X. X. Yi$^{3,\dag}$}

\address{$^1$ Department of Physics, Fuzhou University, Fuzhou 350002, China\\
$^2$ Fujian Key Laboratory of Quantum Information and Quantum Optics (Fuzhou University), Fuzhou 350116, China\\
$^3$ Center for Quantum Sciences and School of Physics, Northeast Normal University, Changchun 130024, China}

\email{*xia-208@163.com} %% email address is required
\email{$\dag$yixx@nenu.edu.cn}



\begin{abstract}
By periodical two-step modulation, we demonstrate that the dynamics of multilevel system can still evolve even in multiple large detunings regime, and provide the effective Hamiltonian (of interest) for this system.
We then illustrate this periodical modulation in quantum state engineering, including achieving direct transition from the ground state to the Rydberg state or
the desired superposition of two Rydberg states without satisfying
two-photon resonance condition,
switching between Rydberg blockade regime and Rydberg antiblockade regime, stimulating distinct atomic transitions by the same laser field, and implementing selective transitions in the same multilevel system.
Particularly, it is robust against perturbation of control parameters.
Another advantage is that the waveform of laser field has simple square-wave
form which is readily implemented in experiments. Thus, it offers us a novel method of quantum state engineering in quantum information processing.
\end{abstract}

%%%%%%%%%%%%%%%%%%%%%%%%%%  body  %%%%%%%%%%%%%%%%%%%%%%%%%%
\section{Introduction}

Perfect coherent control of quantum state is a critical technology in quantum  information processing \cite{scully97,nielsen10}.
The traditional way usually adopts (near-) resonance driving fields.
For instance, with regard to the transition from ground state to Rydberg state, two laser fields should satisfy so-called two-photon resonance condition with the help of intermediate state \cite{schempp15,thaichroen15}.
One disadvantage of resonance condition is that system dynamics seriously suffers from the amplitude noise of driving fields \cite{gardiner99}. Furthermore, the resonance condition in fact is rigorous from experimental viewpoint, since it requires to accurately match driving field frequency with transition frequency. This is also the essential reason why the same laser field cannot stimulate different atomic transitions.
But if this unexplored problem is resolved, i.e., not requiring resonance condition in dynamics any longer, it can afford us a novel avenue to implement coherent manipulations by driving fields with arbitrary frequencies in the same quantum system.

On the other hand, there is increasingly interested in periodic driving systems \cite{kitagawa10,zhou10,rudner13,tong14,piskunow14,chen15,ponte15,iwahori16,shi16,yang17,pagel17,desbuquoisl17}, to a large extent, due to the fact that periodic driving fields provide the possibility of controlling and changing some physical properties of system, such as tunable spin-orbit coupling in ultracold-atom systems \cite{jimenez15,luo16} and the appearance of Floquet Majorana fermions \cite{lindner11,jiang11,liu13,katan13,torres14,benito14}.
For periodic driving systems, they possess well-defined Floquet eigenstates and corresponding eigenvalues (i.e., so-called quasienergies), which can be readily calculated in an extend Hilbert space \cite{shirley65,sambe73}.
By modulating Floquet eigenstates and quasienergies, the dynamical behaviors of periodic driving systems would appear more interesting than that of static systems.
Recently, the periodic driving field has been employed to study the St\"{u}ckelberg interference phenomenon \cite{shevchenko10,huang11,silveri15} and the dynamics of three-level system \cite{kenmoe16}.

To our knowledge, there are several ways such as the chirped adiabatic passage \cite{malinovskaya07,malinovskaya07a,kuznetsova14,malinovskaya17} that can be employed to implement population transfer when two-photon resonance is not satisfied well.
However, it is scarcely ever studied the transition of multilevel system in multiple large detunings (MLDs) regime (i.e., beyond two (multi)-photon resonance).
One of the reasons may be that the analytical solutions of eigenvalues and eigenstates are hard to obtain in multilevel system.
Thus, it is a very fresh attempt of applying periodic driving fields in MLDs regime.
In this work, we demonstrate that quantum state can still evolve by periodical two-step modulation in MLDs regime, however large the detunings are.
Interestingly, the system dynamics would be strikingly different when taking different time intervals of two-step modulation.
Due to the existence of multiple single-photon detunings, some levels can be safely eliminated so that the system is effectively reduced to two-level model.
Particularly, it takes full advantage of large detuning conditions, i.e., the system is robust against perturbation of control parameters and the dynamics is frozen again when removing periodical modulation \cite{scully97}.
We then illustrate its applications in quantum state engineering, including the direct transition from the ground state to the Rydberg state (or the desired superposition of two Rydberg states). The choice of control  parameters is flexibility in Rydberg atom system, e.g., the laser frequencies and modulation periods can be arbitrary in principle.
Other applications are to switch between Rydberg blockade
regime \cite{jaksch00,lukin01,gaetan09,urban09} and Rydberg antiblockade regime \cite{ates07,amthor10,su17}, stimulate different atomic transitions by the same laser field, and implement selective transitions in the same multilevel system.

\section{Two-step modulation in general three-level system}

Consider the toy model that a general three-level atomic system interacts with two laser fields, where the level transition $|k\rangle\leftrightarrow|k+1\rangle$ ($k=1,2$) is coupled by the $k$-th laser field with coupling strength $\Omega_k$.
In the lab frame, the system Hamiltonian reads ($\hbar=1$ hereafter)
\begin{eqnarray}
\mathcal{H}_0=\sum\limits_{k=1}^{3}\omega_k|k\rangle\langle k|+\sum\limits_{k=1}^{2}\Omega_ke^{-i\omega^l_kt}|k\rangle\langle k+1|+H.c.,
\end{eqnarray}
where $\omega_k$ and $\omega^l_k$ are the frequency of level $|k\rangle$ and the $k$-th laser field, respectively. For convenience, it is instructive to adopt interaction picture and the system Hamiltonian becomes
\begin{eqnarray} \label{1s}
H_0=\sum\limits_{k=1}^{2}\Delta_k|k+1\rangle\langle k+1|+\Omega_k|k\rangle\langle k+1|+H.c.,
\end{eqnarray}
where $\Delta_1=\omega_2-\omega_1-\omega^l_1$, $\Delta_2=\omega_3-\omega_1-\omega^l_1-\omega^l_2$.
One of the most attracting points in this work is that there is no restriction with detuning $\Delta_k$ ($k=1,2$), while the two-photon (near-) resonance condition (i.e., $\Delta_2\simeq0$) should be always satisfied during evolution process in previous work.
Without loss of generality, we assume the system is in MLDs regime, i.e., $\Delta_k\gg\Omega_k$, and there are no other restrictions on the values of $\Delta_k$ and $\Omega_k$ ($k=1,2$).

To achieve the evolution operator of system, we need to know the system eigenstates and eigenvalues.
In mathematic, by solving a cubic equation, one can analytically calculate the eigenstates and eigenvalues of Hamiltonian $H_0$ in Eq. (\ref{1s}), then achieve the evolution operator subsequently. However, it is helpless for us in this work due to the intricate expressions of eigenstates and eigenvalues.
In the following, we adopt alternative method to approximatively achieve its expressions. Before elaborating this method, it is instructive to adopt the matrix form of Hamiltonian $H_0$ in the basis $\{|1\rangle,|2\rangle,|3\rangle\}$, which reads
\begin{eqnarray}
H_0=\left(
                   \begin{array}{ccc}
                   0 & \Omega_1 &  0 \\
                   \Omega_1 & \Delta_1 & \Omega_2 \\
                   0 & \Omega_2 & \Delta_2 \\
                   \end{array}
                   \right).
\end{eqnarray}
At first, by imposing the unitary transformation $S=\left(
                   \begin{array}{ccc}
                   1 & 0 &  0 \\
                   0 & \sin\alpha & \cos\alpha \\
                   0 & \cos\alpha & -\sin\alpha \\
                   \end{array}
                   \right)$ on the system, the Hamiltonian becomes
\begin{eqnarray}
\mathbb{H}_0=S^+H_0S=\left(
                   \begin{array}{ccc}
                   0 & \Omega_1\sin\alpha &  \Omega_1\cos\alpha \\
                   \Omega_1\sin\alpha & \xi_1 & 0 \\
                   \Omega_1\cos\alpha & 0 & \xi_2 \\
                   \end{array}
                   \right),
\end{eqnarray}
where $\tan2\alpha=\frac{2\Omega_2}{\Delta_2-\Delta_1}$ and $\xi_{1,2}=\frac{1}{2}(\Delta_1+\Delta_2\mp\sqrt{(\Delta_1-\Delta_2)^2+4\Omega_2^2})$.
Note that the Hilbert space is composed by the basis $\{|1\rangle,|\xi_1\rangle,|\xi_2\rangle\}$ now,
where $|\xi_{1}\rangle=\sin\alpha|2\rangle+\cos\alpha|3\rangle$ and $|\xi_{2}\rangle=\cos\alpha|2\rangle-\sin\alpha|3\rangle$.
When $\Omega_1\ll\{\xi_1,\xi_2,|\xi_1-\xi_2|\}$, according to the standard perturbation theory, the eigenvalues $E_k$ and eigenstates $|E_k\rangle$ ($k=1,2,3$) of $\mathbb{H}_0$ approximatively read
\begin{eqnarray}
E_1&\simeq&-\xi_1x_1^2-\xi_2x_2^2, ~~~~|E_1\rangle\simeq|1\rangle-x_1|\xi_1\rangle- x_2|\xi_2\rangle, \cr
E_2&\simeq&\xi_1+\xi_1x_1^2, ~~~~~~~~~|E_2\rangle\simeq|\xi_1\rangle+x_1|1\rangle,  \cr
E_3&\simeq&\xi_2+\xi_2x_2^2, ~~~~~~~~~|E_3\rangle\simeq|\xi_2\rangle+x_2|1\rangle.
\end{eqnarray}
where $x_1=\frac{\Omega_1\sin\alpha}{\xi_1}\ll1$, $x_2=\frac{\Omega_1\cos\alpha}{\xi_2}\ll1$, and the coefficient of eigenstates is not normalization.
As a result, the evolution operator of system reads
\begin{eqnarray}
\mathcal{U}(t)=e^{-i\mathbb{H}_0t}=\scriptstyle\left(
                   \begin{array}{ccc}
                   1 & x_1(e^{-i\Theta_1}-1) &  x_2(e^{-i\Theta_2}-1) \\
                   x_1(e^{-i\Theta_1}-1) & e^{-i\Theta_1} & x_1x_2 \\
                   x_2(e^{-i\Theta_2}-1) & x_1x_2 & e^{-i\Theta_2} \\
                   \end{array}
                   \right),
\end{eqnarray}
where we have ignored the global phase and the higher-order terms ($\sim x_1^2, x_2^2$), $\Theta_1=(E_2-E_1)t$, and $\Theta_2=(E_3-E_1)t$.
Naturally, the expression of evolution operator in the basis $\{|1\rangle,|2\rangle,|3\rangle\}$ reads
\begin{eqnarray}    \label{18}
&&U(t)=S\mathcal{U}(t)S^+       \cr
&&\simeq\scriptstyle\left(
                   \begin{array}{ccc}
                   \scriptscriptstyle1 & \scriptscriptstyle(e^{-i\Theta_2}-1)x_1\sin\alpha+(e^{-i\Theta_1}-1)x_2\cos\alpha &  \scriptscriptstyle(e^{-i\Theta_2}-1)x_1\cos\alpha+(1-e^{-i\Theta_1})x_2\sin\alpha \\
                   \scriptscriptstyle(e^{-i\Theta_2}-1)x_1\sin\alpha+(e^{-i\Theta_1}-1)x_2\cos\alpha & \scriptscriptstyle e^{-i\Theta_2}\sin^2\alpha+e^{-i\Theta_1}\cos^2\alpha & \scriptscriptstyle(e^{-i\Theta_2}-e^{-i\Theta_1})\cos\alpha\sin\alpha \\
                   \scriptscriptstyle(e^{-i\Theta_2}-1)x_1\cos\alpha+(1-e^{-i\Theta_1})x_2\sin\alpha & \scriptscriptstyle(e^{-i\Theta_2}-e^{-i\Theta_1})\cos\alpha\sin\alpha & \scriptscriptstyle e^{-i\Theta_2}\cos^2\alpha+e^{-i\Theta_1}\sin^2\alpha \\
                   \end{array}
                   \right). \nonumber\\
\end{eqnarray}
It is easily observed from Eq. (\ref{18}) that the populations of each level are almost frozen under
static Hamiltonian $H_0$ in Eq. (\ref{1s}) in MLDs regime.

From the above derivation procedure, we can find that only employing static Hamiltonian $H_0$ is utterly helpless for quantum state engineering. However, The situation is quite different when we adopt periodical two-step modulation of Hamiltonian, i.e.,
\begin{eqnarray}  \label{19}
H(t)=\left \{
\begin{array}{ll}
    H_0=\sum\limits_{k=1}^{2}\Delta_k|k+1\rangle\langle k+1|+\Omega_k|k\rangle\langle k+1|+H.c.,    ~~& t\in[mT,mT+\tau_1), \\
    H_0'=\sum\limits_{k=1}^{2}\Delta_k'|k+1\rangle\langle k+1|+\Omega_k'|k\rangle\langle k+1|+H.c., ~~& t\in[mT+\tau_1,(m+1)T), \\
\end{array}
\right.
\end{eqnarray}
where $T$ is the period of two-step modulation, $m\in \mathbb{N}$. Note that the system is still in MLDs regime if only employing static Hamiltonian $H_0'$.
All control parameters of Hamiltonian $H_0'$ are added the label ``$\prime$'' to distinguish the Hamiltonian $H_0$, and we set $\tau'_1=T-\tau_1$ for brevity hereafter.
Here, the remarkable difference with other periodic works \cite{goldman14,ribeiro17,Claeys17} is that we cannot employ the Baker-Campbell-Hausdorff expansion to calculate the effective Hamiltonian $H_{eff}$, since the driving frequency $\omega=\frac{2\pi}{T}$ is demanded for the same magnitude to the energy gap in this periodic system.
The method we adopt is to directly calculate the evolution operator within a period $T$, then achieve the effective Hamiltonian $H_{eff}$ by definition $U(T)\equiv e^{-iH_{eff}T}$ \cite{eastham73}. The detail derivation process is as follows.
At first, we suppose the time interval $\tau_1^{\scalebox{.5}(\prime\scalebox{.5})}$ satisfying the condition: $\Theta_1^{\scalebox{.5}(\prime\scalebox{.5})}= ({E_2^{\scalebox{.5}(\prime\scalebox{.5})}-E_1^{\scalebox{.5} (\prime\scalebox{.5})}})\tau_1^{\scalebox{.5}(\prime\scalebox{.5})}=(2n+1)\pi$, $n\in \mathbb{N}$, and $\Theta_2^{\scalebox{.5}(\prime\scalebox{.5})}= (E_3^{\scalebox{.5}(\prime\scalebox{.5})}-E_1^{\scalebox{.5}(\prime\scalebox{.5})}) \tau_1^{\scalebox{.5}(\prime\scalebox{.5})}=\phi^{\scalebox{.5}(\prime\scalebox{.5})}$. Obviously, the time interval $\tau_1^{\scalebox{.5}(\prime\scalebox{.5})}$ would be small when the detuning $\Delta_1$ is large in MLDs regime. According to Eq. (\ref{18}), the evolution operator of system in a period $T$ reads,
\begin{eqnarray}
U(T)=U(\tau_1')U(\tau_1)=e^{-iH_0'\tau_1'}e^{-iH_0\tau_1}
    \simeq\left(
                   \begin{array}{ccc}
                   1 & z_1 &  z_2 \\
                   z_4 & 1 & z_3 \\
                   z_5 & z_6 & {e^{-i(\phi+\phi')}} \\
                   \end{array}
                   \right),
\end{eqnarray}
where
\begin{eqnarray}
&&z_1=y_1-y_1'+y_2'y_3, ~~z_2=y_2+e^{-i\phi}y_2'+y_1'y_3, ~~z_3=y_1'y_2-y_3+e^{-i\phi}y_3', \cr
&&z_4=y_1'-y_1+y_2y_3', ~~z_5=y_2'+e^{-i\phi'}y_2+y_1y_3', ~~z_6=y_1y_2'-y_3'+e^{-i\phi'}y_3, \cr
&&y_1^{(\prime)}=(e^{-i\phi^{(\prime)}}-1)x_1^{(\prime)}\sin\alpha^{(\prime)}- 2x_2^{(\prime)}\cos\alpha^{(\prime)}, ~~y_2^{(\prime)}=(e^{-i\phi^{(\prime)}}-1)x_1^{(\prime)}\cos\alpha^{(\prime)}-2x_2^{(\prime)} \sin\alpha^{(\prime)},   \cr
&&y_3^{(\prime)}=(e^{-i\phi^{(\prime)}}+1)\cos\alpha^{(\prime)}\sin\alpha^{(\prime)}. \nonumber
\end{eqnarray}

Note that there are many ways to choose control parameters to perform two-step modulation in principle.
For instance, one can only modulate the detunings $\Delta_k$ and $\Delta_k'$ ($k=1$ or $k=2$) while keeping other control parameter fixed, or one can only modulate the coulping strengths $\Omega_k$ and $\Omega_k'$ while keeping other control parameter fixed, and so on.
Here, we exemplify one of them: $\Delta_k'=\Delta_k$, $\Omega_2'=\Omega_2$, and $\Omega_1'=-\Omega_1$, i.e., employing two-step modulation of the coupling strength $\{\Omega_1,-\Omega_1\}$ (other situations of two-step modulation are demonstrated by numerical simulations in Fig. \ref{fig:02} later).
As a result, the evolution operator in a period $T$ approximatively reads
\begin{eqnarray}  \label{10a}
U(T)\simeq \left(
                   \begin{array}{ccc}
                   \cos(\phi_1)&  -\sin(\phi_1) & 0  \\
                   \sin(\phi_1) & \cos(\phi_1) & 0 \\
                   0 & 0 & e^{-i\varphi_1} \\
                   \end{array}
                   \right),
\end{eqnarray}
where $\phi_1=\frac{4\Omega_1}{\xi_1}\sin^2\alpha+\frac{4\Omega_1}{\xi_2}\cos^2\alpha$. By reversely solving the matrix equation $U(T)=e^{-iH_{eff}T}$, the effective Hamiltonian of two-step modulation reads
\begin{eqnarray}  \label{4a}
H_{eff}=\Omega_{eff}|1\rangle\langle2|+\Omega_{eff}^*|2\rangle\langle1|,
\end{eqnarray}
where $\Omega_{eff}=\frac{i\phi_1}{T}$.
Therefore, the dynamics of system by two-step modulation is reduced to ``Rabi oscillation'' between $|1\rangle$ and $|2\rangle$ with the effective ``Rabi frequency'' $\Omega_{eff}$, without exciting the high level $|3\rangle$.

\begin{figure}[htbp]
\centering
\includegraphics[scale=0.36]{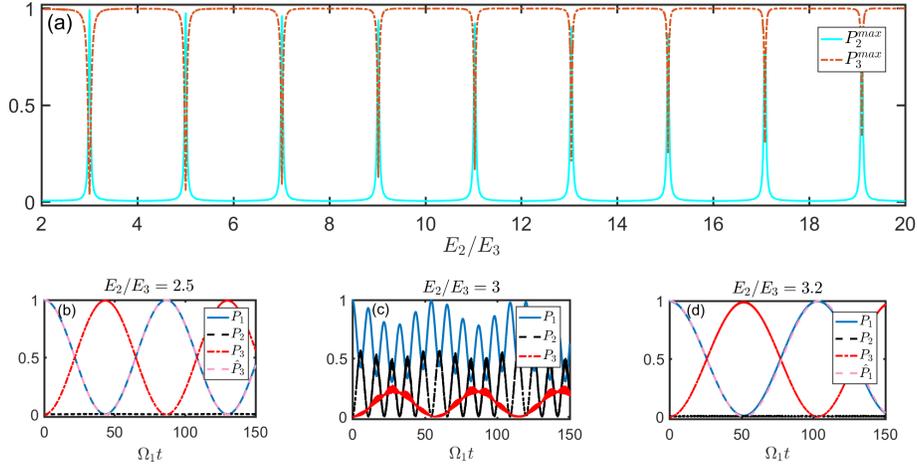}
\caption{ (a) The maximum population $P_k^{max}$ of level $|k\rangle$ ($k=2,3$) during the whole evolution versus $E_2/E_3$ by periodically modulating the coupling strength $\{\Omega_1,\Omega_1'\}$, where the initial state is $|1\rangle$ and $\Delta_1/\Omega_1=60$, $\Omega_2^{(\prime)}/\Omega_1=2$, $\Omega'_1=-\Omega_1$, $\Delta_k'=\Delta_k$ ($k=1,2$), $\tau_1^{(\prime)}=\frac{\pi}{E_3^{(\prime)}-E_1^{(\prime)}}$. (b-d) The time evolution of populations $P_k$ ($k=1,2,3$) of each levels with different $E_2/E_3$. Except for pink-dash line, all system dynamics are simulated by using Hamiltonian (\ref{19}).}  \label{fig:01}
\end{figure}

\begin{figure}[hbtp]
\centering
\includegraphics[scale=0.3]{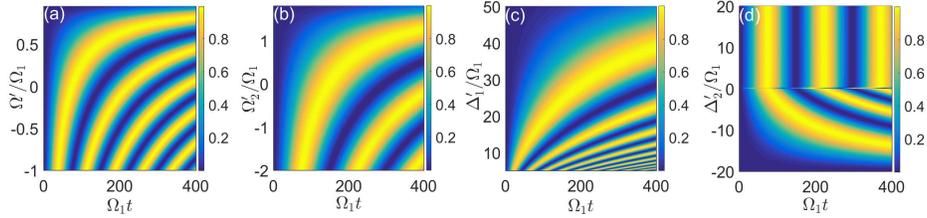}
\caption{(a) The population $P_3$ versus evolution time and coupling strength $\Omega_1'$ under two-step modulation of the coupling strength $\{\Omega_1,\Omega_1'\}$, where $\Omega_2/\Omega_1=2$, $\Delta_1/\Omega_1=60$, $\Delta_2/\Omega_1=30$, $\Omega_2'=\Omega_2$, $\Delta_k'=\Delta_k$ ($k=1,2$), $\tau_1^{(\prime)}=\frac{\pi}{E_3^{(\prime)}-E_1^{(\prime)}}$. (b) The population $P_3$ versus evolution time and coupling strength $\Omega_2'$ under two-step modulation of the coupling strength $\{\Omega_2,\Omega_2'\}$, where $\Omega_2/\Omega_1=2$, $\Delta_1/\Omega_1=60$, $\Delta_2/\Omega_1=-20$, $\Omega_1'=\Omega_1$, $\Delta_k'=\Delta_k$ ($k=1,2$), $\tau_1^{(\prime)}=\frac{\pi}{E_3^{(\prime)}-E_1^{(\prime)}}$. (c) The population $P_3$ versus evolution time and detuning $\Delta_1'$ under two-step modulation of the detuning $\{\Delta_1,\Delta_1'\}$, where $\Omega_2/\Omega_1=2$, $\Delta_1/\Omega_1=60$, $\Delta_2/\Omega_1=-30$, $\Omega_k'=\Omega_k$ ($k=1,2$), $\Delta_2'=\Delta_2$, $\tau_1^{(\prime)}=\frac{\pi}{E_3^{(\prime)}-E_1^{(\prime)}}$. (d) The population $P_3$ versus evolution time and detuning $\Delta_2'$ under two-step modulation of the detuning $\{\Delta_2,\Delta_2'\}$, where $\Omega_2/\Omega_1=2$, $\Delta_1/\Omega_1=60$, $\Delta_2/\Omega_1=-20$, $\Omega_k'=\Omega_k$ ($k=1,2$), $\Delta_1'=\Delta_1$, $\tau_1^{(\prime)}=\frac{\pi}{E_3^{(\prime)}-E_1^{(\prime)}}$. }  \label{fig:02}
\end{figure}

On the other hand, if we set the time interval $\tau_1^{\scalebox{.5}(\prime\scalebox{.5})}$ satisfying the condition: $\Theta_2^{\scalebox{.5}(\prime\scalebox{.5})}= ({E_3^{\scalebox{.5}(\prime\scalebox{.5})}-E_1^{\scalebox{.5} (\prime\scalebox{.5})}})\tau_1^{\scalebox{.5}(\prime\scalebox{.5})}=(2n+1)\pi$, $n\in \mathbb{N}$, and $\Theta_1^{\scalebox{.5}(\prime\scalebox{.5})}= (E_2^{\scalebox{.5}(\prime\scalebox{.5})}-E_1^{\scalebox{.5}(\prime\scalebox{.5})}) \tau_1^{\scalebox{.5}(\prime\scalebox{.5})}=\phi^{\scalebox{.5}(\prime\scalebox{.5})}$,
similar above derivation process, the evolution operator in a period $T$ reads,
\begin{eqnarray}
U(T)&=&U(\tau_1')U(\tau_1)=e^{-iH_0'\tau_1'}e^{-iH_0\tau_1} \cr
    &\simeq&\scriptstyle\left(
                   \begin{array}{ccc}
                   1 & z_1 &  z_2 \\
                   -z_1 & {e^{-i2\phi}\cos^2\alpha+\sin^2\alpha} & z_3 \\
                   -z_2 & z_3 & {e^{-i2\phi}\sin^2\alpha+\cos^2\alpha} \\
                   \end{array}
                   \right),
\end{eqnarray}
where
$z_1=4x_1\sin\alpha+e^{-2i\phi}(e^{i\phi}-1)^2 x_2\cos\alpha$,
$z_2=4x_1\cos\alpha-e^{-2i\phi}(e^{i\phi}-1)^2 x_2\sin\alpha$,
and $z_3=(1-e^{-i2\phi})\cos\alpha\sin\alpha$.
When $e^{i\phi}=-1$, i.e., $E_2=(2n+3) E_3, n\in\mathbb{N}$,
we find that $z_1=4x_1\sin\alpha+4x_2\cos\alpha$,
$z_2=4x_1\cos\alpha-4x_2\sin\alpha\simeq0$, and $z_3=0$.
Thus, it is invalid for quantum state engineering between $|1\rangle$ and $|3\rangle$ by two-step modulation.
However, if $e^{i\phi}=1$, i.e., $E_2=2(n+1) E_3,n\in\mathbb{N}$,
we find that $z_1=4x_1\sin\alpha\simeq0$,
$z_2=4x_1\cos\alpha$, and $z_3=0$.
In this case, by reversely solving the matrix equation $U(T)=e^{-iH_{eff}T}$, we achieve the effective Hamiltonian of two-step modulation:
\begin{eqnarray} \label{10}
H_{eff}'=\Omega_{eff}'|1\rangle\langle3|+\Omega_{eff}^{\prime*}|3\rangle\langle1|,
\end{eqnarray}
where $\Omega_{eff}'=\frac{i\phi_1'}{T}$, $\phi_1'=2\frac{\Omega_1}{\xi_1}\sin2\alpha$.
Therefore, the dynamics of system by two-step modulation is reduced to ``Rabi oscillation'' between $|1\rangle$ and $|3\rangle$ with the effective ``Rabi frequency'' $\Omega_{eff}'$, without exciting the intermediate level $|2\rangle$.
In fact, except for the narrow spike region: $E_2/E_3\simeq(2n+3), n\in\mathbb{N}$, the effective Hamiltonian of two-step modulation can be always represented by Eq. (\ref{10}), which is verified by numerical simulations in Fig. \ref{fig:01} (the maximum population $P_3^{max}$ of level $|3\rangle$ can mainly reach unit except for this narrow spike region).
In particular, the pink-dash line in Fig. \ref{fig:01}(b) and Fig. \ref{fig:01}(d) also represent the time evolution of population of level $|1\rangle$, but they are plotted by the effective Hamiltonian $H_{eff}'$ in Eq. (\ref{10}). The highly consistent between blue-solid line and pink-dash line demonstrates that the effective Hamiltonian  $H_{eff}'$ is valid for describing the system  dynamics by two-step modulation.
%Hence, two-step modulation can be employed to quantum state engineering in MLDs regime and it is interesting to find that choosing different time interval $\tau_1^{\scalebox{.5}(\prime\scalebox{.5})}$ of two-step modulation would lead to different system evolutions.

Finally, it is worth mentioning that two-step modulation of other control parameters can also be used to implement quantum state engineering  in MLDs regime.
Since the derivation process is similar to the process of periodically modulating coupling strength $\{\Omega_1,-\Omega_1\}$, we just present the numerical results in periodically modulating other control parameters by choosing the time interval $\tau_1^{\scalebox{.5}(\prime\scalebox{.5})}= \frac{\pi}{E_3^{\scalebox{.5}(\prime\scalebox{.5})}-E_1^{\scalebox{.5}(\prime\scalebox{.5})}}$,
which implement the transition between $|1\rangle$ and $|3\rangle$.
Figures \ref{fig:02}(a)-\ref{fig:02}(d) respectively denote the time evolution of population $P_3$ by two-step modulation of coupling strength $\{\Omega_1,\Omega_1'\}$, coupling strength $\{\Omega_2,\Omega_2'\}$, detuning $\{\Delta_1,\Delta_1'\}$, and detuning $\{\Delta_2,\Delta_2'\}$, witnessing the feasibility of two-step modulation in three-level system.
To be more specific, an inspection of Fig. \ref{fig:02}(a) shows that the period of Rabi oscillation is long when increasing the coupling strength $\Omega_1'$. In particular, the transition process become showly when
the coupling strength $\Omega_1'$ approaches to $\Omega_1$. This consequence is not surprising since two-step modulation reverts to one-step modulation if $\Omega_1'=\Omega_1$ and the system dynamics is frozen in this regime.
Similar results are also achieved for two-step modulation of coupling strength $\{\Omega_2, \Omega_2'\}$ and detuning $\{\Delta_1, \Delta_1'\}$, as shown in Figs. \ref{fig:02}(b)-\ref{fig:02}(c).
However, as shown in Fig. \ref{fig:02}(d), it is quite different from the two-step modulation of detuning $\{\Delta_2, \Delta_2'\}$, where the period of Rabi oscillation remains unchange if $\Delta_2'>0$ and increases with the decreasing of $\Delta_2'$ if $\Delta_2'<0$.

\section{Applications in Rydberg systems}

\begin{figure}[tbp]
\centering
\includegraphics[scale=0.7,bb=110 570 540 750]{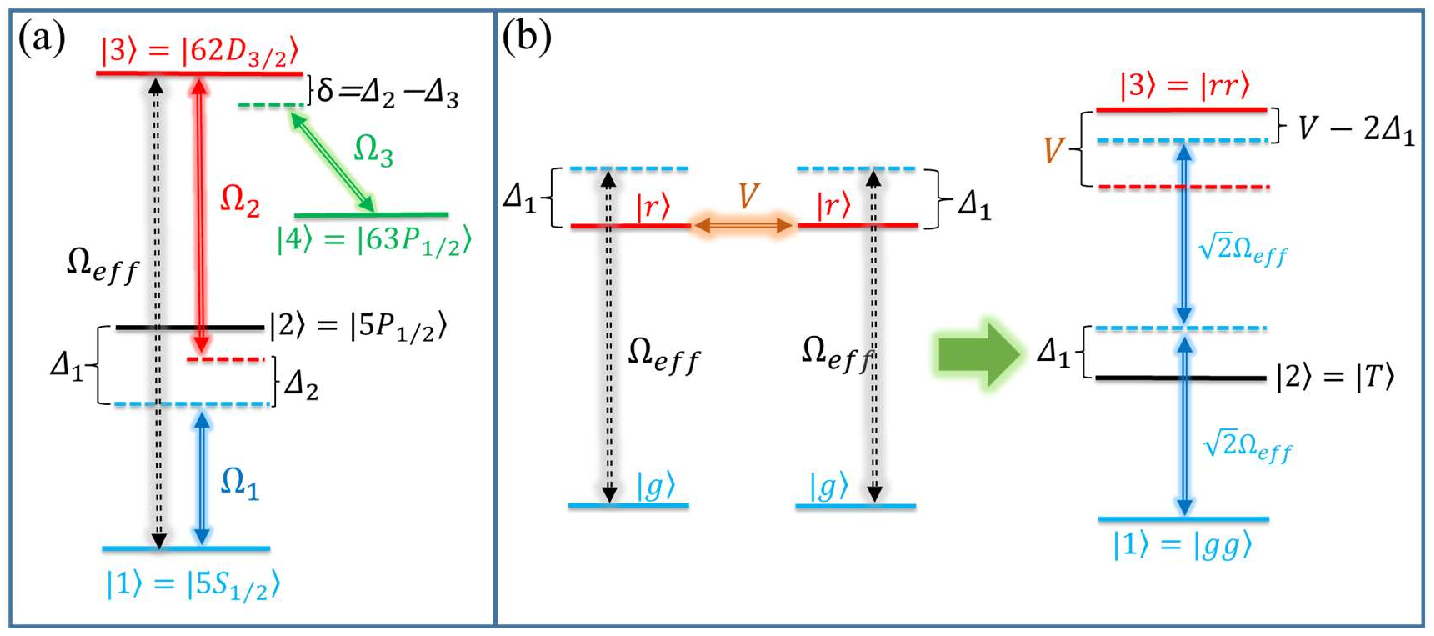}
\caption{(a) The structure of Rydberg atom coupled by two laser fields and a microwave field. (b) The structure of two identical atoms coupled by laser fields and Rydberg-Rydberg interaction.}  \label{sfig:01}
\end{figure}

\begin{figure}[b]
\centering
\includegraphics[scale=0.34]{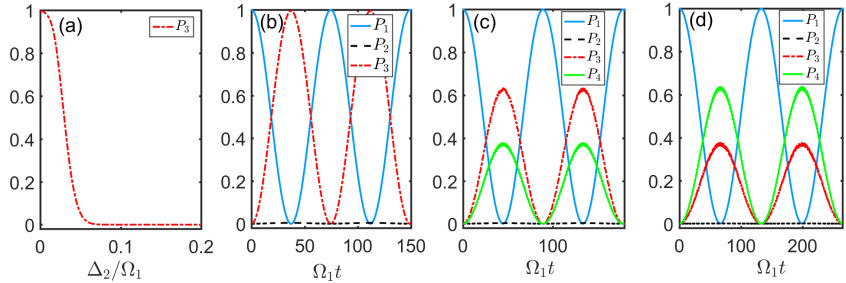}
\caption{(a) The population $P_3$ of level $|3\rangle$ versus $\Delta_2$ in STIRAP, where $\Omega_1(t)=\Omega_1e^{-\frac{(t-\tau)^2}{\tau^2}}$, $\Omega_2(t)=\Omega_1e^{-\frac{t^2}{\tau^2}}$, $\tau=200$, and $\Delta_1/\Omega_1=30$.
(b) The time evolution of populations $P_k$ ($k=1,2,3$) of each levels by periodically modulating coupling strength $\{\Omega_1,\Omega'_1\}$  in three-level system, where $\Omega'_1/\Omega_1=-1$, $\Delta_1^{(\prime)}/\Omega_1=60$, $\Delta_2^{(\prime)}/\Omega_1=30$, $\Omega_2^{(\prime)}/\Omega_1=2$, $\tau_1^{(\prime)}=\frac{\pi}{E_3^{(\prime)}-E_1^{(\prime)}}$.
(c-d) The time evolution of populations $P_k$ ($k=1,2,3,4$) of each levels by periodically modulating coupling strength $\{\Omega_1,\Omega'_1\}$  in four-level system, where $\Omega'_1/\Omega_1=-1$, $\Delta_1^{(\prime)}/\Omega_1=60$, $\Delta_2^{(\prime)}/\Omega_1=30$, $\Omega_2^{(\prime)}/\Omega_1=2$, $\Delta_3^{(\prime)}/\Omega_1=28.8$ , $\Omega_3^{(\prime)}/\Omega_1=2$. The time interval satisfies $\tau_1^{(\prime)}=\frac{\pi}{E_3^{(\prime)}-E_1^{(\prime)}}$ in panel (c) while the time interval satisfies  $\tau_1^{(\prime)}=\frac{\pi}{E_4^{(\prime)}-E_1^{(\prime)}}$ in panel (d). All system dynamics are simulated by Hamiltonian (\ref{19}).
}  \label{fig:02a}
\end{figure}

In this section, we take Rydberg state transitions as first example. As shown in Fig. \ref{sfig:01}(a), a Rydberg atom is coupled by two laser fields, where the coupling strengths are $\Omega_1$ and $\Omega_2$ respectively.
The states $|5S_{1/2}\rangle$, $|5P_{1/2}\rangle$, and $|62D_{3/2}\rangle$ correspond to the levels $|1\rangle$, $|2\rangle$, and $|3\rangle$, respectively.
To stimulate the Rydberg state $|62D_{3/2}\rangle$ from the ground state $|5S_{1/2}\rangle$, one usually adopts stimulated Raman adiabatic passage (STIRAP) and it requires two-photon resonance condition (i.e., $\Delta_2=0$) between two laser fields in experiments \cite{tresp16,thaicharoen17}. In other words, two laser frequencies must be chosen specially, e.g., the wavelengths of laser fields are 795 nm and 474 nm in Fig. \ref{sfig:01}(a)  \cite{barredo15}, respectively.
But what happen when the two-photon resonance condition cannot be satisfied well? Clearly, as shown in Fig. \ref{fig:02a}(a), the population of Rydberg state $|62D_{3/2}\rangle$ sharply drops with the increase of $\Delta_2/\Omega_1$, verifying that STIRAP is invalid even the detuning $\Delta_2$ is small (also cf. the ``narrow spikes'' of Fig. 13 in \cite{bergmann98}).
To solve this issue, we employ two-step modulation, so that one can safely ignore two-photon resonance condition. The resulting benefits is that two laser frequencies can be chosen arbitrarily now.

At the start we can arbitrarily choose two laser frequencies to ensure the system in MLDs regime. Note that different laser frequencies only affect the period of two-step modulation.
Then we set the system in the ground state $|5S_{1/2}\rangle$ initially, and the population of each levels are frozen due to MLDs regime. Next we periodically modulate coupling strength to implement ``Rabi oscillation'' between ground state $|5S_{1/2}\rangle$ (=$|1\rangle$) and Rydberg state $|62D_{3/2}\rangle$ (=$|3\rangle$).
Figure \ref{fig:02a}(b) demonstrates the system dynamics when the time interval satisfies $\tau_1^{\scalebox{.5}(\prime\scalebox{.5})}=\frac{\pi}{E_3^{\scalebox{.5}(\prime\scalebox{.5})} -E_1^{\scalebox{.5}(\prime\scalebox{.5})}}$ in two-step modulation of coupling strength $\{\Omega_1,\Omega'_1\}$, and the Rydberg state is achieved at specific time $t_s=37.2/\Omega_1$.
In practice, this process is easily realized by modulating the phase of laser field with square-wave generator to produce $\pi$-phase difference \cite{cho08,chen10,keating16,zhang17,shi17}.
After achieving the Rydberg state, we need to remove phase modulator since the populations are frozen again in MLDs regime.
Note that if two different coupling strengths are adopted in two-step modulation, it needs an attenuation modulator rather than phase modulator.

The second application of two-step modulation is to prepare desired superposition of two Rydberg states: $|\psi\rangle=\cos\vartheta|62D_{3/2}\rangle+\sin\vartheta|63P_{1/2}\rangle$, which is exploited for fast Rydberg quantum gate \cite{shi17}.
In the experiment \cite{barredo15}, $|\psi\rangle$ is achieved by several operation steps, including switching off and on the laser fields and the microwave field in sequence. In particular this experiment also requires $\Delta_2=0$ and $\delta=0$ in Fig. \ref{sfig:01}(a). However, the two-step modulation offers a quite simple way to achieve this goal, which only requires periodically modulating the coupling strength $\{\Omega_1,\Omega'_1\}$.
In Figs. \ref{fig:02a}(c)-\ref{fig:02a}(d), one directly drives the ground state to the desired superposition of two Rydberg states by choosing specific time $t_s$, where the angle $\vartheta$ is determined by the coupling strength $\Omega_3$ and detuning $\delta$, i.e., $\vartheta\simeq\frac{1}{2}\tan^{-1}\frac{2\Omega_3}{\delta}$. To be more specific, if we choose the coupling strengths $\Omega_1=100$MHz, $\Omega_2=\Omega_3=200$MHz, the detunings $\Delta_1=6$GHz, $\Delta_2=3$GHz, $\delta=120$MHz, the period of phase modulator would be $T=2$ns. Hence, the superposition of two Rydberg states is achieved at time $t_s=446$ns by two-step modulation of coupling strength $\{\Omega_1,-\Omega_1\}$.

\begin{figure}[b]
\centering
\includegraphics[scale=0.34]{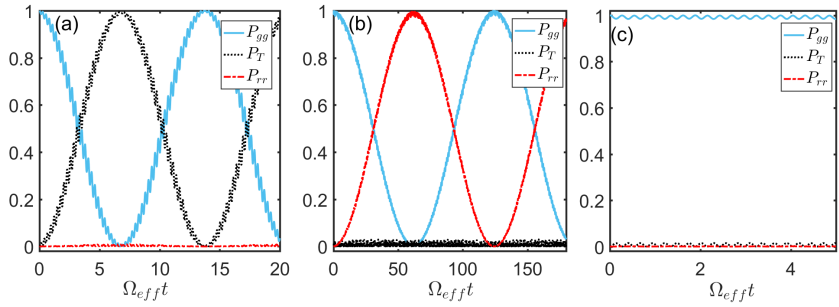}
\caption{ The time evolution of populations $P_m$ ($m=gg, T, rr$) of each levels by periodically modulating coupling strength $\{\Omega_{eff},\Omega'_{eff}\}$, where $\Omega_{eff}'=0.5\Omega_{eff}$, $\Delta_1/\Omega_{eff}=-23$, $V/\Omega_{eff}=39$. (a) $\tau_1^{(\prime)}=\frac{\pi}{E_{T}^{(\prime)}-E_{gg}^{(\prime)}}$, (b) $\tau_1^{(\prime)}=\frac{\pi}{E^{(\prime)}_{rr}-E_{gg}^{(\prime)}}$. (c) The time evolution of populations $P_m$ ($m=gg, T, rr$) of each levels without two-step modulation, $\Delta_1/\Omega_{eff}=-23$, $V/\Omega_{eff}=39$.}  \label{fig:s4}
\end{figure}

Another application of two-step modulation in Rydberg atoms is that we can control the transition between Rydberg blockade regime and Rydberg antiblockade regime. As shown in Fig. \ref{sfig:01}(b), two identical atoms have ground state $|g\rangle$ and Rydberg state $|r\rangle$. The $|g\rangle\leftrightarrow|r\rangle$ transition is coupled by laser fields with the effective coupling strength $\Omega_{eff}$ and the detuning $\Delta_1$. At the same time, there exists Rydberg-Rydberg interaction between two atoms, where the interaction strength is $V$. Thus, the system Hamiltonian reads
\begin{eqnarray}
\mathcal{H}_0=\Omega_{eff}e^{i\Delta_1t}(|g\rangle_{11}\langle r|\otimes\mathbb{I}_{2}+ \mathbb{I}_{1}\otimes|g\rangle_{22}\langle r|+H.c.)+V|rr\rangle\langle rr|,
\end{eqnarray}
where $\mathbb{I}_k$ denotes the identity operator of the $k$-th atom ($k=1,2$), and $|mn\rangle$ is the abbreviation of $|m\rangle_1|n\rangle_2$ ($m,n=g,r$). In the rotation frame defined by $R=e^{-i(\Delta_1|T\rangle\langle T|+2\Delta_1|rr\rangle\langle rr|)t}$, the system Hamiltonian becomes
\begin{eqnarray}
H_0=-\Delta_1|T\rangle\langle T|+(V-2\Delta_1)|rr\rangle\langle rr|+\sqrt{2}\Omega_{eff}(|T\rangle\langle gg|+|T\rangle\langle rr|+H.c.),
\end{eqnarray}
where $|T\rangle=\frac{1}{\sqrt{2}}(|gr\rangle+|rg\rangle)$.
When $\Delta_1=0$, the system is in the Rydberg blockade regime, i.e., the doubly excited Rydberg state $|rr\rangle$ cannot be stimulated from $|T\rangle$ directly. However, when $\Delta_1=\frac{V}{2}$, the system is in the Rydberg antiblockade regime, i.e., the doubly excited Rydberg states $|rr\rangle$ can be stimulated from $|gg\rangle$. In other words, it requires different laser fields to make the system in Rydberg blockade regime or Rydberg antiblockade regime.
Here, we demonstrate that the Rydberg blockade regime and Rydberg antiblockade regime can be switched by merely regulating the period of two-step modulation in the same laser fields.
Figure \ref{fig:s4} presents the system dynamics by periodically modulating coupling strength $\{\Omega_{eff},\Omega'_{eff}\}$.
As shown in Fig. \ref{fig:s4}(a), when we choose the time interval $\tau_1^{\scalebox{.5}(\prime\scalebox{.5})}$ satisfy: $\tau_1^{\scalebox{.5}(\prime\scalebox{.5})}=\frac{\pi}{E_{T}^{\scalebox{.5}(\prime\scalebox{.5})} -E_{gg}^{\scalebox{.5}(\prime\scalebox{.5})}}$, it emerges Rabi oscillation between $|gg\rangle$ and $|T\rangle$ and hinders the transition to $|rr\rangle$. That is, the system is in the Rydberg blockade regime. However, as shown in Fig. \ref{fig:s4}(b), when we choose the time interval $\tau_1^{\scalebox{.5}(\prime\scalebox{.5})}$ satisfy: $\tau_1^{\scalebox{.5}(\prime\scalebox{.5})}=\frac{\pi}{E_{rr}^{\scalebox{.5}(\prime\scalebox{.5})} -E_{gg}^{\scalebox{.5}(\prime\scalebox{.5})}}$, it emerges Rabi oscillation between $|gg\rangle$ and $|rr\rangle$, which means the system is in the Rydberg antiblockade regime. Note that the slight oscillation in the populations $P_m$ can be effectively eliminated by decreasing the time interval $\tau_1^{\scalebox{.5}(\prime\scalebox{.5})}$ of two-step modulation.
When removing the two-step modulation, the system is in neither Rydberg blockade regime nor Rydberg antiblockade regime, as shown in Fig. \ref{fig:s4}(c).
Therefore, we can regulate the time interval $\tau_1^{\scalebox{.5}(\prime\scalebox{.5})}$ of two-step modulation to determine the system in Rydberg blockade regime or Rydberg antiblockade regime, or neither of them, which does not require different laser fields now.

\section{Implementation of selective transitions in multilevel system}

\begin{figure}[b]
\centering
\includegraphics[scale=0.76,bb=110 570 540 750]{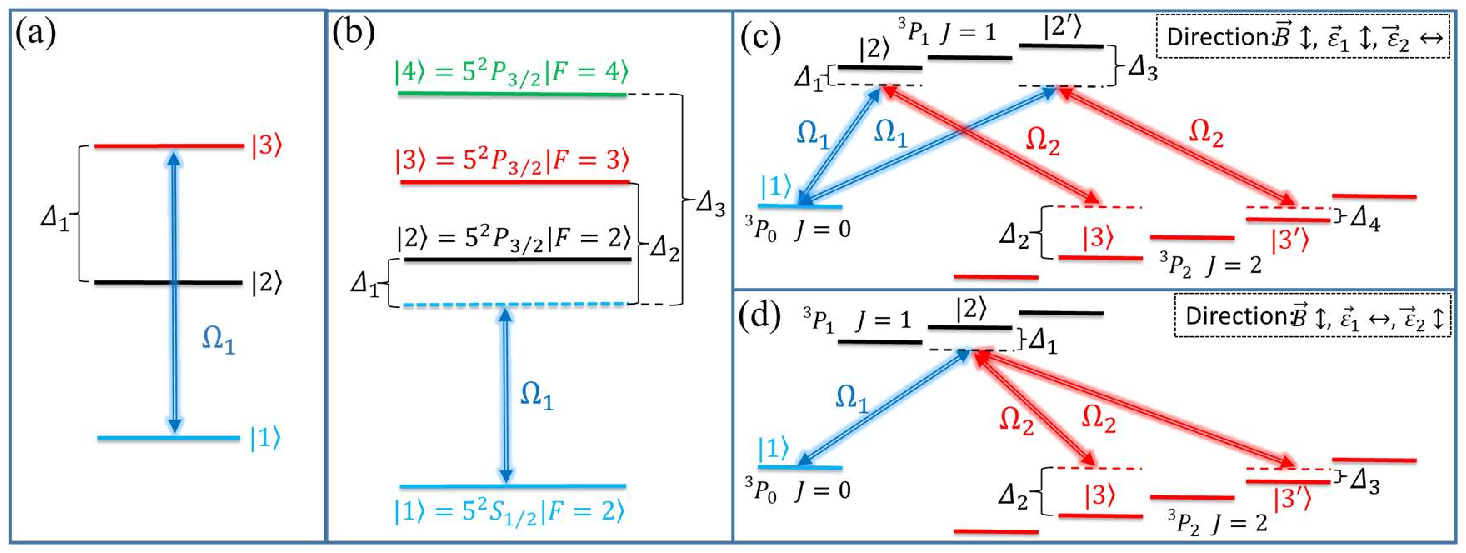}
\caption{(a) The structure of three-level system coupled by a laser field, where the detuning $\Delta_1$ exactly matches with the transition frequency $\omega_{23}$. (b) The structure of Rubidium atom driven by single laser field with large detunings. (c-d) The structure of Ne$^*$ atom coupled by laser fields $\vec{\varepsilon}_1$ and $\vec{\varepsilon}_2$, where the degeneracy of sublevels are removed by magnetic field $\vec{B}$.}  \label{fig:01a}
\end{figure}

\begin{figure}[htbp]
\centering
\includegraphics[scale=0.36]{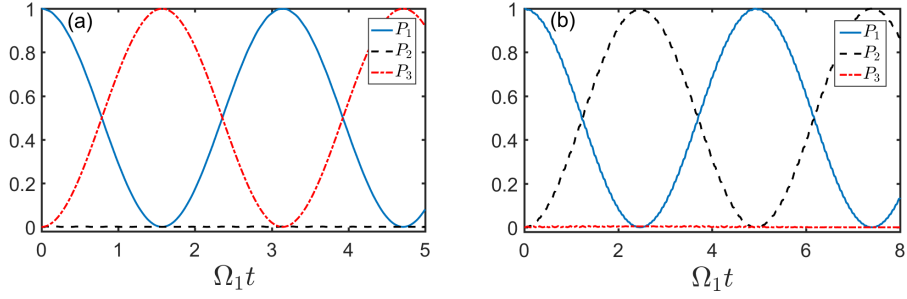}
\caption{ The time evolution of populations $P_m$ ($m=1,2,3$) of each levels (a)without, (b)with,  periodically modulating coupling strength $\{\Omega_1,\Omega_1'\}$, where $\Omega_1'=-\Omega_1$, $\Delta_1/\Omega_1=-48$, $\tau_1=\frac{\pi}{E_2-E_1}$, $\tau_1'=\frac{\pi}{E_3'-E_1'}$. }  \label{fig:032a}
\end{figure}

In most situations, multilevel nature of quantum system needs to be considered, since it may make invalid for the hypothesis that the laser fields only interact with two levels. We first investigate a counterintuitive phenomenon in two-step modulation. The physical model consists of a three-level system coupled by single laser field with coupling strength $\Omega_1$, where we artificially add large detuning $\Delta_1$ that exactly matches with the transition frequency $\omega_{23}$, as shown in Fig. \ref{fig:01a}(a).
In the rotating frame, the Hamiltonian reads
\begin{eqnarray}
H_0=\Delta_1|2\rangle\langle2|+\Omega_1|1\rangle\langle2|+\Omega_1|1\rangle\langle3|+H.c.
\end{eqnarray}
Naturally, if we do not employ two-step modulation, the laser field would only stimulate the $|1\rangle\leftrightarrow|3\rangle$ transition due to the resonance condition, as shown in Fig. \ref{fig:032a}(a). However, we verify in Fig. \ref{fig:032a}(b) that, regardless of resonance condition, the laser field would stimulate the $|1\rangle\leftrightarrow|2\rangle$ transition by two-step modulation, sharply suppressing the transition to the level $|3\rangle$.
As a result, even though the laser field resonantly drives the transition between $|1\rangle$ and $|3\rangle$, the $|1\rangle\leftrightarrow|3\rangle$ transition still cannot occur by two-step modulation.

\begin{figure}[htbp]
\centering
\includegraphics[scale=0.33]{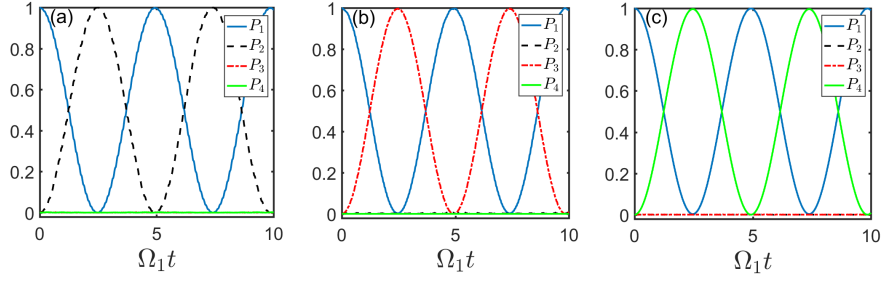}
\caption{ The time evolution of populations $P_m$ ($m=1,2,3$) of each levels by periodically modulating coupling strength $\{\Omega_1,\Omega'_1\}$ with different time interval $\tau_1^{(\prime)}$, where $\Omega'_1=-\Omega_1$, $\Delta_1/\Omega_1=30$, $\Delta_2/\Omega_1=53$, $\Delta_3/\Omega_1=100$. (a) $\tau_1^{(\prime)}=\frac{\pi}{E_2^{(\prime)}-E_1^{(\prime)}}$, (b) $\tau_1^{(\prime)}=\frac{\pi}{E_3^{(\prime)}-E_1^{(\prime)}}$, (c) $\tau_1^{(\prime)}=\frac{\pi}{E_4^{(\prime)}-E_1^{(\prime)}}$.}  \label{fig:031a}
\end{figure}

This counterintuitive phenomenon can be applied to implement different atomic transitions with single laser field by artificially adding MLDs in multilevel systems.
As shown in Fig. \ref{fig:01a}(b), suppose that the multilevel atom is coupled by a single laser field, which drives $|1\rangle\leftrightarrow|2\rangle$ transition with large detuning $\Delta_1$, drives the $|1\rangle\leftrightarrow|3\rangle$ transition with large detuning $\Delta_2$, and drives the $|1\rangle\leftrightarrow|4\rangle$ transition with large detuning $\Delta_3$.
In the rotating frame, the Hamiltonian reads
\begin{eqnarray}
H_0=\sum_{k=2}^{4}\Delta_{k-1}|k\rangle\langle k|+\Omega_1|1\rangle\langle k|+\Omega_1|k\rangle\langle 1|.
\end{eqnarray}
There is no doubt that the laser field cannot drive any transition without any modulations in system due to MLDs condition.
However, this situation is changed by two-step modulation.
Figure \ref{fig:031a} demonstrates different atomic transition is achieved by two-step modulation of the same laser field.
As a concrete example, with regard to the Rubidium 85 D$_2$ transition hyperfine structure \cite{daniel08}, one can stimulate the $5^2S_{1/2}|F=2\rangle\leftrightarrow5^2P_{3/2}|F=2\rangle$ transition with the time interval  $\tau_1^{\scalebox{.5}(\prime\scalebox{.5})}= \frac{\pi}{E_2^{\scalebox{.5}(\prime\scalebox{.5})}-E_1^{\scalebox{.5}(\prime\scalebox{.5})}}$ in Fig. \ref{fig:031a}(a), stimulate the $5^2S_{1/2}|F=2\rangle\leftrightarrow5^2P_{3/2}|F=3\rangle$ transition with the time interval $\tau_1^{\scalebox{.5}(\prime\scalebox{.5})}= \frac{\pi}{E_3^{\scalebox{.5}(\prime\scalebox{.5})}-E_1^{\scalebox{.5}(\prime\scalebox{.5})}}$ in Fig. \ref{fig:031a}(b), or stimulate the $5^2S_{1/2}|F=2\rangle\leftrightarrow5^2P_{3/2}|F=4\rangle$ transition with the time interval $\tau_1^{\scalebox{.5}(\prime\scalebox{.5})}= \frac{\pi}{E_4^{\scalebox{.5}(\prime\scalebox{.5})}-E_1^{\scalebox{.5}(\prime\scalebox{.5})}}$ in Fig. \ref{fig:031a}(c) by two-modulation of coupling strength $\{\Omega_1,-\Omega_1\}$. In other words, different periods of two-step modulation would determine different atomic transitions in the same system.

\begin{figure}[htbp]
\centering
\includegraphics[scale=0.35]{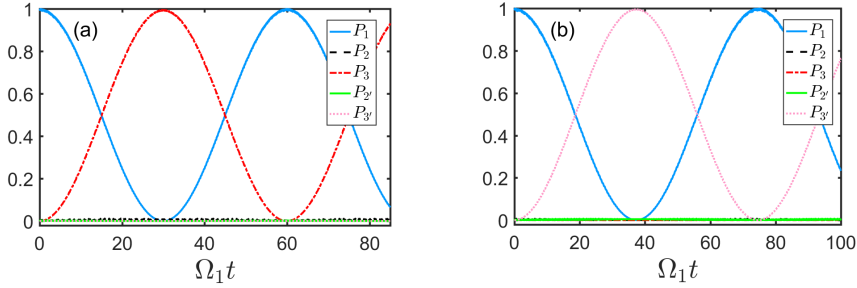}
\caption{ The time evolution of populations $P_m$ ($m=1,2,3,2',3'$) of each levels by periodically modulating coupling strength $\{\Omega_1,\Omega'_1\}$ in five-level system, where $\Omega'_1=-\Omega_1$, $\Omega_2/\Omega_1=2$, $\Delta_1/\Omega_1=33$, $\Delta_2/\Omega_1=9$, $\Delta_3/\Omega_1=36$, $\Delta_4/\Omega_1=6$. (c) $\tau_1^{(\prime)}=\frac{\pi}{E_3^{(\prime)}-E_1^{(\prime)}}$, (d) $\tau_1^{(\prime)}=\frac{\pi}{E^{(\prime)}_{3'}-E_1^{(\prime)}}$.}  \label{fig:031}
\end{figure}

Other applications of two-step modulation can be found in the Ne$^*$ atom system. As shown in Fig. {\ref{fig:01a}}(c),  when the laser field $\vec{\varepsilon}_1$ is parallel to the magnetic field $\vec{B}$ while the laser field $\vec{\varepsilon}_2$ is perpendicular to the magnetic field $\vec{B}$, it would be reduced to five-level system.
To be specific, the laser field $\vec{\varepsilon}_1$ drives the transitions $|1\rangle\leftrightarrow|2\rangle$ and $|1\rangle\leftrightarrow|2'\rangle$ with the coupling strength $\Omega_1$, while the laser field $\vec{\varepsilon}_2$ drives the transitions $|2\rangle\leftrightarrow|3\rangle$ and $|2'\rangle\leftrightarrow|3'\rangle$ with the coupling strength $\Omega_2$.
In the rotating frame, the system Hamiltonian reads
\begin{eqnarray}
H_0&=&\Delta_1|2\rangle\langle2|+\Delta_3|2'\rangle\langle2'| +(\Delta_1+\Delta_2)|3\rangle\langle3| +(\Delta_3+\Delta_4)|3'\rangle\langle3'|  \cr
&&+  \Omega_1|1\rangle\langle2|+\Omega_1|1\rangle\langle2'| +\Omega_2|2\rangle\langle3|+\Omega_2|2'\rangle\langle3'|+H.c.
\end{eqnarray}
With the help of two-step modulation, the transition paths are
selective by choosing different time interval $\tau_1^{\scalebox{.5}(\prime\scalebox{.5})}$ in this system. For instance, if $\tau_1^{\scalebox{.5}(\prime\scalebox{.5})}= \frac{\pi}{E_3^{\scalebox{.5}(\prime\scalebox{.5})}-E_1^{\scalebox{.5}(\prime\scalebox{.5})}}$, as shown in Fig. \ref{fig:031}(a), we achieve the level $|3\rangle$ through transition path `$|1\rangle-|2\rangle-|3\rangle$'. If $\tau_1^{\scalebox{.5}(\prime\scalebox{.5})}= \frac{\pi}{E_{3'}^{\scalebox{.5}(\prime\scalebox{.5})}-E_1^{\scalebox{.5}(\prime\scalebox{.5})}}$, as shown in Fig. \ref{fig:031}(b), we achieve the level $|3'\rangle$ through transition path `$|1\rangle-|2'\rangle-|3'\rangle$'. In practice, we just change the period of square-wave on phase modulator to realize two different transition paths.

\begin{figure}[htbp]
\centering
\includegraphics[scale=0.35]{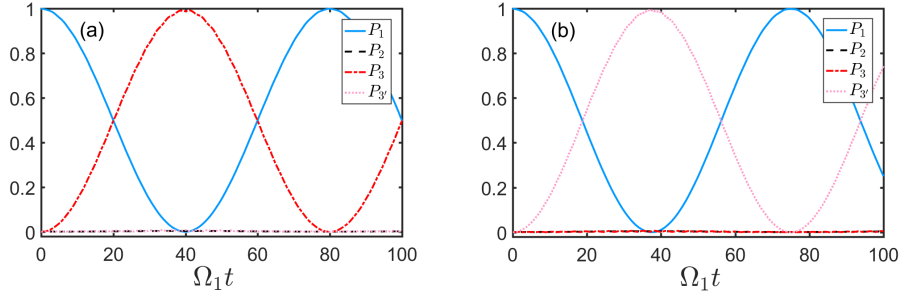}
\caption{ The time evolution of populations $P_k$ ($k=1,2,3,3'$) of each levels by periodically modulating coupling strength $\{\Omega_1,\Omega'_1\}$, where $\Omega_2/\Omega_1=2$, $\Delta_1/\Omega_1=60$, $\Delta_2/\Omega_1=30$, $\Delta_3/\Omega_1=28$, $\Omega'_1=-\Omega_1$, $\Omega'_2=\Omega_2$, $\Delta_k'=\Delta_k, (k=1,2,3)$. (a) $\tau_1^{(\prime)}=\frac{(2n+1)\pi}{E^{ (\prime)}_{3}-E_1^{(\prime)}}$. (b) $\tau_1^{(\prime)}=\frac{(2n+1)\pi}{E^{ (\prime)}_{3'}-E_1^{(\prime)}}$. }  \label{fig:05}
\end{figure}

However, if the laser field $\vec{\varepsilon}_1$ is perpendicular to the magnetic field $\vec{B}$ while the laser field $\vec{\varepsilon}_2$ is parallel to the magnetic field $\vec{B}$, only four levels are coupled by laser fields in this system, as shown in Fig. \ref{fig:01a}(d). To be specific, the laser field $\vec{\varepsilon}_1$ drives the $|1\rangle\leftrightarrow|2\rangle$ transition with Rabi frequency $\Omega_1$ and detuning $\Delta_1$.
The laser field $\vec{\varepsilon}_2$ drives the $|2\rangle\leftrightarrow|3\rangle$ and $|2\rangle\leftrightarrow|3'\rangle$ transition with Rabi frequency $\Omega_2$ and detunings $(\Delta_1+\Delta_2)$ and $(\Delta_1+\Delta_3)$, respectively.
Thus, the system Hamiltonian reads
\begin{eqnarray}
H_{0}=\Delta_1|2\rangle\langle2|+\Delta_2|3\rangle\langle3|+\Delta_3|3'\rangle\langle3'| +\Omega_1|1\rangle\langle2|+\Omega_2|2\rangle\langle3|+\Omega_2|2\rangle\langle3'|+H.c.
\end{eqnarray}
We also adopt two-step modulation of coupling strength $\{\Omega_1,\Omega_1'\}$ while all other physical parameters remain unchanged.
If the time interval $\tau_1^{\scalebox{.5}(\prime\scalebox{.5})}$ satisfies $\tau_1^{\scalebox{.5}(\prime\scalebox{.5})}=\frac{(2n+1)\pi}{E_3^{\scalebox{.5}(\prime\scalebox{.5})} -E_1^{\scalebox{.5}(\prime\scalebox{.5})}}$, Rabi oscillation between levels $|1\rangle$ and $|3\rangle$ occurs, as shown in Fig. \ref{fig:05}(a). However, if the time interval $\tau_1^{\scalebox{.5}(\prime\scalebox{.5})}$ satisfies $\tau_1^{\scalebox{.5}(\prime\scalebox{.5})}=\frac{(2n+1)\pi}{E^{\scalebox{.5}(\prime\scalebox{.5})}_{3'} -E_1^{\scalebox{.5}(\prime\scalebox{.5})}}$, Rabi oscillation between levels $|1\rangle$ and $|3'\rangle$ occurs, as shown in Fig. \ref{fig:05}(b).
That is, we can control selective transition by choosing different periods of two-step modulation.

\section{Robust against perturbation of control parameters}

\begin{figure}[htbp]
\centering
\includegraphics[scale=0.3]{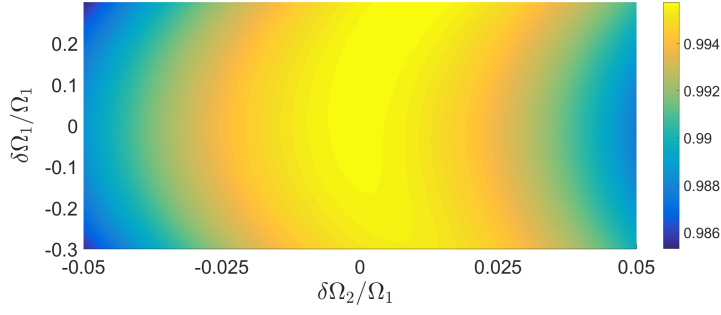}
\caption{ The population $P_3$ of Rydberg state versus the perturbations $\delta\Omega_1$ and $\delta\Omega_2$
in two-step modulation of coupling strength $\{\Omega_1,\Omega'_1\}$, where $\Omega'_1/\Omega_1=-1$, $\Delta_1^{(\prime)}/\Omega_1=60$, $\Delta_2^{(\prime)}/\Omega_1=30$, $\Omega_2^{(\prime)}/\Omega_1=2$, $\tau_1^{(\prime)}=\frac{\pi}{E_3^{(\prime)}-E_1^{(\prime)}}$. }  \label{fig:11}
\end{figure}

Until now, we have studied the physical model in ideal case, i.e., all control parameters are known accurately. However, in practice, the control parameters of system unavoidably exist some unknown perturbations so that it would affect the evolution process. As a result, it is very essential to examine whether or not the two-step modulation is valid in the presence of perturbations.
At first, we exemplify the influence of perturbations in laser fields on Rydberg state preparation by periodically modulating coupling strength $\{\Omega_1,\Omega_1'\}$, where the system Hamiltonian is rewritten as
\begin{eqnarray}
H(t)=\left \{
\begin{array}{ll}
    \scriptstyle\Delta_1|2\rangle\langle 2|+\Delta_2|3\rangle\langle 3|+(\Omega_1+\delta\Omega_1)|1\rangle\langle 2|+(\Omega_2+\delta\Omega_2)|2\rangle\langle 3|+H.c.,    & \scriptstyle t\in[mT,mT+\tau), \\
    \scriptstyle\Delta_1|2\rangle\langle 2|+\Delta_2|3\rangle\langle 3|+(\Omega_1'+\delta\Omega_1)|1\rangle\langle 2|+(\Omega_2+\delta\Omega_2)|2\rangle\langle 3|+H.c., & \scriptstyle t\in[mT+\tau,(m+1)T). \\
\end{array}
\right.
\end{eqnarray}
$\delta\Omega_k$ ($k=1,2$) denotes the strength of unknown perturbations in coupling strength $\Omega_k$.
Figure \ref{fig:11} represents the population $P_3$ of Rydberg state versus perturbations $\delta\Omega_1$ and $\delta\Omega_2$ at the evolution time $t_s=37.2/\Omega_1$.
We can find that the population of Rydberg state is still high ($\geq0.988$) even though there exist the perturbation ($|\delta\Omega_2/\Omega_1|\leq0.05$) in the coupling strength $\Omega_2$.
Particularly, it is almost immune to the perturbation $\delta\Omega_1$, which stems from the fact that the energy gap of system is hardly affected by the perturbation $\delta\Omega_1$ when periodically modulating coupling strength $\Omega_1$.

\begin{figure}[htbp]
\centering
\includegraphics[scale=0.36]{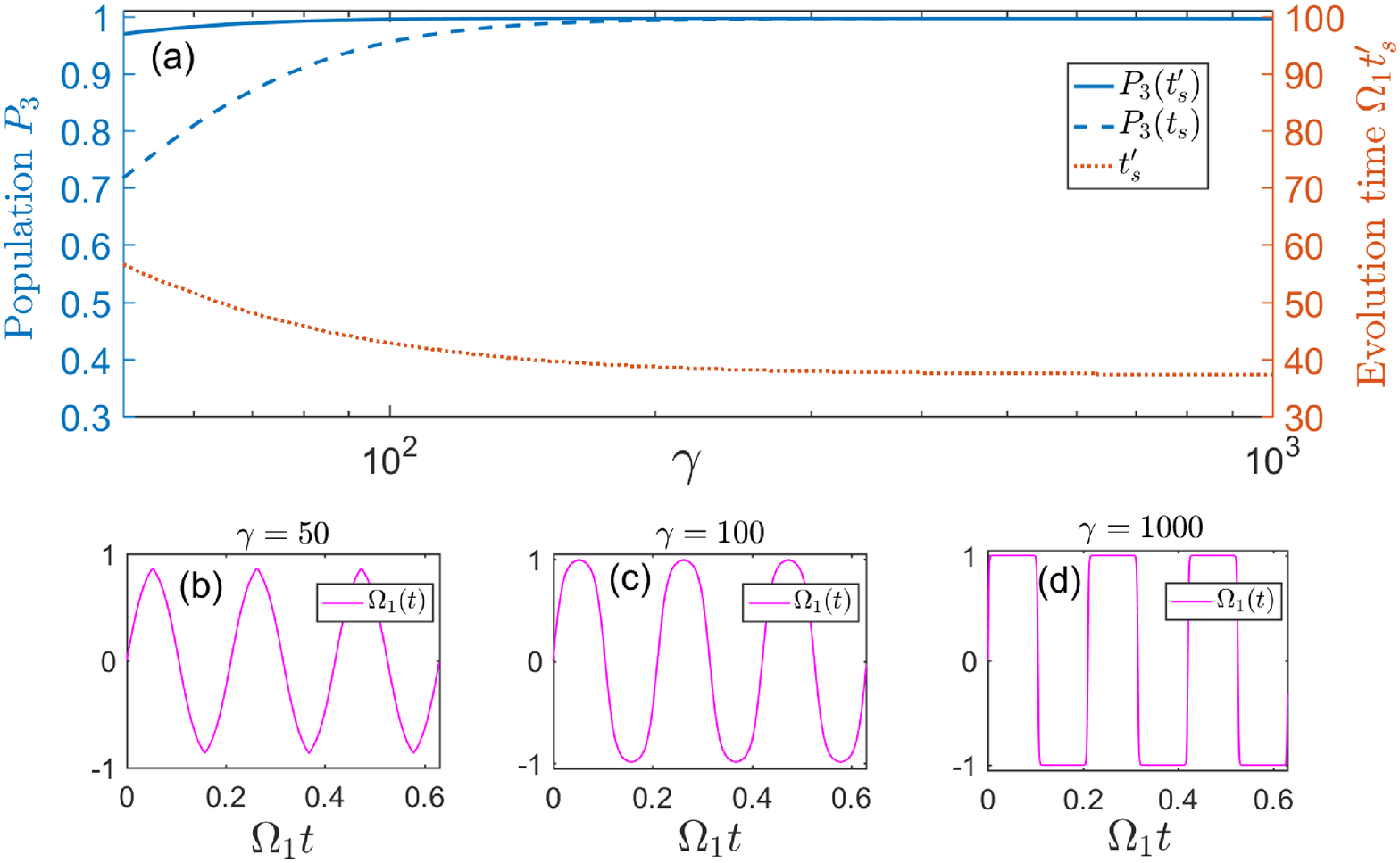}
\caption{ (a) The population $P_3$ of Rydberg state (the left-blue vertical axis) and the evolution time $t'_s$ achieving the maximum population $P_3$ (the right-orange vertical axis) as a function of $\gamma$ in two-step modulation of coupling strength $\{\Omega_1,\Omega'_1\}$, where $\Omega'_1/\Omega_1=-1$, $\Delta_1^{(\prime)}/\Omega_1=60$, $\Delta_2^{(\prime)}/\Omega_1=30$, $\Omega_2^{(\prime)}/\Omega_1=2$, $\tau_1^{(\prime)}=\frac{\pi}{E_3^{(\prime)}-E_1^{(\prime)}}$. (b-d) The shapes of coupling strength $\Omega_1(t)$ with different $\gamma$. (b) $\gamma=50$. (c) $\gamma=100$. (d) $\gamma=1000$.
}  \label{fig:12}
\end{figure}

On the other hand, the square-wave produced by waveform generator might not be perfect in experiments. In the following, we study the influence of the square-wave deformation of coupling strength $\Omega_1(t)$ on Rydberg state preparation, where the expression now reads
\begin{eqnarray}  \label{21}
\Omega_1(t)=\left \{
\begin{array}{lll}
    \Omega'_1+\frac{\Omega_1-\Omega'_1}{1+e^{-\gamma\textrm{mod}(t/T)}}, &
    \textrm{mod}(t/T)<\frac{\tau}{2},\\
    \Omega'_1+\frac{\Omega_1-\Omega'_1}{1+e^{\gamma[\textrm{mod}(t/T)-\tau]}}, & \frac{\tau}{2}\leq\textrm{mod}(t/T)\leq T-\frac{\tau'}{2}, \\
    \Omega'_1+\frac{\Omega_1-\Omega'_1}{1+e^{-\gamma[\textrm{mod}(t/T)-T]}}, & \textrm{mod}(t/T)>T-\frac{\tau'}{2}.\\
\end{array}
\right.
\end{eqnarray}
The dimensionless parameter $\gamma$ represents the hardness of square-wave. Specifically, the shape of coupling strength $\Omega_1(t)$ in Eq. (\ref{21}) gradually approaches to square-wave when $\gamma$ is large, and it becomes perfect square-wave if $\gamma\rightarrow\infty$.
As illustrations, Figs. \ref{fig:12}(b)-\ref{fig:12}(d) represent some concrete shapes of coupling strength $\Omega_1(t)$ with different $\gamma$.
In Fig. \ref{fig:12}(a), the blue-dash line denotes the population $P_3$ of Rydberg state as a function of $\gamma$ at evolution time $t_s=37.2/\Omega_1$.
One observes that the population of Rydberg state maintains a relatively high value for most $\gamma$, i.e., the deformation of square-wave has little influence on Rydberg state preparation.
Note that only when the square-wave is heavy deformation, e.g., $\gamma=50$ in Fig. \ref{fig:12}(b), it would affect the preparation process of Rydberg state. In fact, this shape is not square-wave any more, and the amplitude of coupling strength $\Omega_1(t)$ cannot reach unit.
Nevertheless, the heavy deformation problem can be effectively solved by properly shifting the evolution time of two-step modulation.
In Fig. \ref{fig:12}(a), the blue-solid line denotes the maximum population $P_3$ of Rydberg state as a function of $\gamma$ during the evolution process, and the orange-dot line denotes the evolution time $t_s'$ when reaching the maximum population $P_3(t_s')$.
We can easily find that the population $P_3$ of Rydberg state is almost unchanged at evolution time $t'_s$ even though there exists heavy deformation in square-wave.

\section{Conclusion}

We have studied a fantastic phenomenon that quantum states can still evolve by periodic modulation even in MLDs regime.
The purpose of artificially adding MLDs is to take full advantage of large detuning conditions and freeze system dynamics when removing periodic modulation.
By regulating the time interval and the period of two-step modulation, we have demonstrated that the multilevel system can be equivalent into distinct two-level systems.
In particular, the atomic system does not need to satisfy resonance condition and the laser frequencies can be chosen arbitrarily in two-step modulation.
Additionally, this modulation is robust against perturbation of control parameters and is easily implemented in experiments due to simple square-pulse form.

For its applications, we have shown: achieving direct transition from the ground state to the Rydberg state or the desired superposition of two Rydberg states; switching between Rydberg blockade regime and Rydberg antiblockade regime; stimulating distinct atomic transitions by the same laser field; implementing selective transitions in multilevel system.
Recently, the transition of two Rydberg states is controlled by using an addressing beam to produce  detuning in experiments \cite{leseleuc17}. We find that this process can also be controlled by two-step modulation of microwave fields without requiring extra addressing beam.
In a word, this periodic modulation would offer us a powerful tool for quantum state engineering as well as implementing a variety of high-fidelity quantum logic gates.

\section*{Funding}

National Natural Science Foundation of China (11575045, 11674060, 11747011, 11805036, 11534002, 11775048, 61475033); Major State Basic Research Development Program of China (2012CB921601); Fund of Fujian Education Department (JAT170086); Natural Science Foundation of Fujian Province of China (2018J01413).
\\

\end{document}